\documentclass[a4paper,11pt]{article}
\usepackage{pos}

\usepackage{url}
\usepackage{fancybox}
\usepackage[english]{babel}
\usepackage{times}
\usepackage{graphicx}
\usepackage{psfrag}
\usepackage{subfigure}

\usepackage{tabulary}
\usepackage{comment}

\usepackage{comment}
\definecolor{darkred}{rgb}{0.4,0.0,0.0}
\definecolor{darkbred}{rgb}{0.7,0.0,0.0}
\definecolor{darkgreen}{rgb}{0.0,0.4,0.0}
\definecolor{darkblue}{rgb}{0.0,0.0,0.4}
\definecolor{darkmagenta}{rgb}{0.55, 0.0, 0.55}
\definecolor{bole}{rgb}{0.47, 0.27, 0.23}
\definecolor{LightBlue}{rgb}{0.5,0.5,1}

\usepackage{braket}
\usepackage[T1]{fontenc}
\usepackage{amsmath}
\usepackage{amsthm}
\usepackage{url}
\usepackage[utf8]{inputenc}
\usepackage{dcolumn}% eqnarray table columns on decimal point
\usepackage{bm}% bold math
\usepackage{url} 
\usepackage{grffile}
\usepackage{xcolor}
\usepackage{booktabs}
\usepackage{verbatim}

\usepackage{listings}
\newcommand{\be}{\begin{equation}}
\newcommand{\ee}{\end{equation}}
\newcommand{\beq}{\begin{equation}}
\newcommand{\eeq}{\end{equation}}
\newcommand{\bea}{\begin{eqnarray}}
\newcommand{\eea}{\end{eqnarray}}

\newcommand{\bie}{\begin{itemize}}

\newcommand{\eie}{\end{itemize} }
\newcommand{\ben}{ \begin{enumerate}}
\newcommand{\een}{\end{enumerate} }
\newcommand{\bbl}{\begin{block}{} \begin{center}}
\newcommand{\ebl}{\end{center} \end{block}}
\newcommand{\bref}{\begin{flushright} \begin{tiny}}
\newcommand{\eref}{\end{tiny} \end{flushright}}
\newcommand{\sbf}{\boldsymbol}

\usepackage{amsmath}
\usepackage{amssymb}
\usepackage{braket}
\usepackage{array,multirow}
\usepackage{tikz}
\usepackage{dsfont}
\usetikzlibrary{snakes}
\usetikzlibrary{math}
\usetikzlibrary{calc}

\tikzset{decoration={snake,amplitude=0.5mm,segment length=2mm,
		post length=0mm,pre length=0mm}}
\newcommand{\midarrow}{\tikz \draw[-stealth] (0,0) -- +(0,.1);}
\tikzset{every path/.style={line width=0.02 cm}}

\newcommand{\f}[1]{\textbf{#1}}

\newcommand{\BBdisconnected}[5] % {scale}{xoffset}{yoffset}{quark flavour 1}{quark flavour 2}
{
	\begin{tikzpicture}[baseline={([yshift=-.7ex]current bounding box.center)}]
		\draw[color=#4] (0,#1*1.3) -- node {\midarrow}(0,0);
		\draw[decorate, color=#5] (0,#1*1.3) to [bend left=45] (0,0);
		\draw[color=#4] (#1,1.3*#1) -- node {\midarrow} (#1,0);
		\draw[decorate, color=#5] (#1,0) to [bend left=45] (#1,#1*1.3);
		\node at (0.3*#1,1.2*#1) {#2};
		\node at (0.65*#1,1.2*#1) {#3};
	\end{tikzpicture}
}

\newcommand{\BBconnected}[5] % {scale}{xoffset}{yoffset}{quark flavour 1}{quark flavour 2}
{
	\begin{tikzpicture}[baseline={([yshift=-.65ex]current bounding box.center)}]
		\draw[color=#4] (0,#1*1.3) -- node {\midarrow} (0,0);
		\draw[decorate, color=#5] (0,#1*1.3) to (#1,0);
		\draw[color=#4] (#1,1.3*#1) -- node {\midarrow} (#1,0);
		\draw[decorate, color=#5] (0,0) to  (#1,#1*1.3);
		\node at (0.3*#1,1.2*#1) {#2};
		\node at (0.65*#1,1.2*#1) {#3};
	\end{tikzpicture}
}

\newcommand{\Bmeson}[5] % {scale}{xoffset}{yoffset}{quark flavour 1}{quark flavour 2}
{
	\begin{tikzpicture}[baseline={([yshift=-.65ex]current bounding box.center)}]
		\draw[color=#4] (0,0) -- node {\midarrow}(0,#1*1.3);
		\draw[decorate, color=#5] (0,#1*1.3) to [bend left=45] (0,0);
		\node at (0.3*#1,1.2*#1) {#2};
	\end{tikzpicture}
}

\newcommand{\gtapprox}{\raisebox{-0.5ex}{$\,\stackrel{>}{\scriptstyle\sim}\,$}}
\newcommand{\ltapprox}{\raisebox{-0.5ex}{$\,\stackrel{<}{\scriptstyle\sim}\,$}}

\graphicspath{{figures}}

\title{%
	Antistatic-antistatic $\bar Q \bar Q qq$ potentials for $u$, $d$ and $s$ light quarks from lattice QCD 
}
\author[1]{Pedro Bicudo}
\author[2]{Marina Krstic Marinkovic}
\author[3]{Lasse Müller}
\author[3,4]{Marc Wagner}

\affiliation[1]{CeFEMA and Physics department, Instituto Superior Técnico, Av.\ Rovisco Pais, 1049 Lisboa, Portugal}
\affiliation[2]{Institut fur Theoretische Physik, Wolfgang-Pauli-Stra{\ss}e 27, ETH Zurich, 8093 Zurich, Switzerland}
\affiliation[3]{Goethe-Universit\"at Frankfurt am Main, Institut f\"ur Theoretische Physik, Max-von-Laue-Stra{\ss}e 1, D-60438 Frankfurt am Main, Germany}
\affiliation[4]{Helmholtz Research Academy Hesse for FAIR, Campus Riedberg, Max-von-Laue-Stra{\ss}e 12, D-60438 Frankfurt am Main, Germany} 

\emailAdd{ bicudo@tecnico.ulisboa.pt}
\emailAdd{ marinama@ethz.ch}
\emailAdd{ lmueller@itp.uni-frankfurt.de} 
\emailAdd{ mwagner@itp.uni-frankfurt.de}

\abstract{
We report on our ongoing lattice QCD computation of antistatic-antistatic potentials in the presence of two light quarks using the CLS $N_f=2$ gauge configurations and the OpenQ*D codebase. We utilize a set of 16 creation operators, corresponding to 8 sectors characterized by angular momentum and parity quantum numbers for light quarks $qq = (ud - du) / \sqrt{2}$ (isospin $0$), $qq \in \{ uu , (ud + du) / \sqrt{2}, dd \}$ (isospin $1$) and $qq \in \{ us , ds \}$ (isospin $1/2$ and strangeness $-1$). We improve on previous work by considering a large number of off-axis separations of the static antiquarks and by using tree-level improvement. The resulting potentials provide vague indication for one-pion exchange at $\bar Q \bar Q$ separations $r \gtapprox 0.5 \, \text{fm}$.
}

\begin{document}
%%%%%%%%%%%%%%%%%%%%%%%%%%%%%%%%%%%%%%%%%%%%%%%%%%%%%%%%%%%%%%%%%%%%%%%%%%%%%

\maketitle

%FFFFFFFFFFFFFFFFFFFFFFFFFFFFFFFFFFFFFFFFFFFFFFFFFFFFFFFFFFFFFFFFFFFFFFFFFFFFFFFFFFFFFFFFFFFFFFFFFF
%FFFFFFFFFFFFFFFFFFFFFFFFFFFFFFFFFFFFFFFFFFFFFFFFFFFFFFFFFFFFFFFFFFFFFFFFFFFFFFFFFFFFFFFFFFFFFFFFFF
%FFFFFFFFFFFFFFFFFFFFFFFFFFFFFFFFFFFFFFFFFFFFFFFFFFFFFFFFFFFFFFFFFFFFFFFFFFFFFFFFFFFFFFFFFFFFFFFFFF

\section{Motivation}

Using the $\bar Q \bar Q q q$ lattice QCD potential data from Ref.\ \cite{Wagner:2010ad} and inspired by $T_{cc}$ and $T_{bb}$ tetraquarks proposed since 1981 (see e.g.\ Ref.\ \cite{Ader:1981db}), we applied a Coulomb-like screened potential fit in Ref.\ \cite{Bicudo:2012qt} to predict binding energies of possibly existing heavy-light tetraquarks from a Schrödinger equation. For $T_{bb}$ with $I(J^P) = 0(1^+)$ we found $E_B \approx -50 \, \text{MeV}$. Extending the lattice QCD data and carrying out a chiral extrapolation in Ref.\ \cite{Bicudo:2015vta,Bicudo:2015kna} led to $E_B \approx -90 \, \text{MeV}$. However, heavy quark spin effects, which were included in Ref.\ \cite{Bicudo:2016ooe}, decrease the binding energy again to $E_B \approx -60 \, \text{MeV}$.

One of the aims of this long term project is to generate high precision lattice QCD potentials, which can be used in Born-Oppenheimer approaches to reliably predict masses of antiheavy-antiheavy-light-light tetrquarks (for a recent proposal of a very advanced and complete Born-Oppenheimer effective theory see Ref.\ \cite{Berwein:2024ztx}). This could resolve a long standing tension with lattice studies of the $T_{bb}$ tetraquark based on NRQCD (see e.g.\ the recent works \cite{Aoki:2023nzp,Hudspith:2023loy,Alexandrou:2024iwi,Colquhoun:2024jzh}), which find somewhat stronger binding, $E_B \ltapprox -100 \, \text{MeV}$.  

Moreover, at large $\bar Q \bar Q$ separations the $\bar Q \bar Q q q$ system is composed of two static-light mesons, each with isospin $1/2$ and light quark spin $1/2$. These are the same quantum numbers as those of a nucleon $N \in \{ p , n \}$. Because of this, our potentials are expected to be qualitatively similar to the nucleon-nucleon ($N$-$N$) interaction, for instance discussed in Ref.\ \cite{Stoks:1994wp} and fundamental to nuclear physics. Thus, meson exchange may produce small bumps at intermediate and large separations, with opposite sign compared to the main short distance parts. At large separations we even expect a dominating one-pion exchange (OPE) potential with its characteristic tensor structure (see e.g.\ Ref.\ \cite{Beane:2001bc}).

%FFFFFFFFFFFFFFFFFFFFFFFFFFFFFFFFFFFFFFFFFFFFFFFFFFFFFFFFFFFFFFFFFFFFFFFFFFFFFFFFFFFFFFFFFFFFFFFFFF
%FFFFFFFFFFFFFFFFFFFFFFFFFFFFFFFFFFFFFFFFFFFFFFFFFFFFFFFFFFFFFFFFFFFFFFFFFFFFFFFFFFFFFFFFFFFFFFFFFF
%FFFFFFFFFFFFFFFFFFFFFFFFFFFFFFFFFFFFFFFFFFFFFFFFFFFFFFFFFFFFFFFFFFFFFFFFFFFFFFFFFFFFFFFFFFFFFFFFFF

\section{Lattice setup and results}

We performed computations on three CLS ensembles featuring $N_f=2$ $O(a)$ improved Wilson fermions. The ensembles are denoted as A5, G8 and N6 (see Refs.\ \cite{Fritzsch:2012wq,Engel:2014eea}), which differ in the lattice spacing ($a = 0.0755 \, \text{fm}, 0.0658 \, \text{fm}, 0.0486 \, \text{fm}$) and the pion mass ($m_\pi = 331 \, \text{MeV} , 185 \, \text{MeV} , 340 \, \text{MeV}$),  and keep the lattice volume in the range $2.3-4.2 \mathrm{fm}$.  We utilized the openQ*D codebase \cite{Campos:2019kgw}. In addition to using techniques some of us already applied in an independent previous lattice QCD computation of $\bar Q \bar Q q q$ potentials \cite{Bicudo:2015kna} (e.g.\ stochastic timeslice propagators, various smearing techniques), we now also computed off axis separations and replaced the lattice separation by a tree-level improved separation, $\mathbf{r}_\text{lat} \to \mathbf{r}_\text{impr} = 4 \pi  a / G(\mathbf{r} / a)$, where $G$ is the scalar lattice propagator associated with our gauge action.

Our tetraquark interpolating operators are
\begin{eqnarray}
\mathcal{O}_{BB}^{q^{(1)} q^{(2)},\Gamma}(\f{r}_1,\f{r}_2) = (\mathcal{C} \Gamma)_{AB} (\mathcal{C}\tilde{\Gamma})_{CD} \Big(\bar Q^a_C(\f{r}_1) q^{(1),a}_A(\f{r}_1)\Big) \Big(\bar Q^b_D(\f{r}_2) q^{(2),b}_B(\f{r}_2)\Big) ,
\label{eqn:creation_operator_bbud}
\end{eqnarray}
where $q^{(1)} q^{(2)} = (ud - du) / \sqrt{2}$ for isospin $I = 0$, $q^{(1)} q^{(2)} \in \{ uu , (ud + du) / \sqrt{2}, dd \}$ for $I = 1$ and $q^{(1)} q^{(2)} \in \{ us , ds \}$ for $I = 1/2$ and strangeness $-1$. The static quark spins are coupled by the $4 \times 4$ matrix $\tilde \Gamma$ and are essentially irrelevant. The light quark spins are coupled by the $4 \times 4$ matrix $\Gamma$, which provides 16 linearly independent operators for each flavor sector. We compute correlation functions
\begin{eqnarray}
\mathcal{C}_{BB}^{q^{(1)} q^{(2)},\Gamma}(\f{r},t) = \Big\langle \Big(\mathcal{O}_{BB}^{q^{(1)} q^{(2)},\Gamma}(\f{r}_1,\f{r}_2;t)\Big)^\dagger \mathcal{O}_{BB}^{q^{(1)} q^{(2)},\Gamma}(\f{r}_1,\f{r}_2;0) \Big\rangle ,
\end{eqnarray}
where $\f{r} = \f{r}_2 - \f{r}_1$. Before we extract $\bar Q \bar Q q q$ potentials $V^{q^{(1)} q^{(2)},\Gamma}_{BB}(\f{r})$, we subtract twice the static-light meson mass by dividing by the squared correlation function of the lightest static light meson,
\begin{eqnarray}
\frac{\mathcal{C}_{BB}^{q^{(1)} q^{(2)},\Gamma}(\f{r},t)}{(\mathcal{C}_B(t))^2} \propto_{t \to \infty} \exp\Big(-V^{q^{(1)} q^{(2)},\Gamma}_{BB}(\f{r}) t\Big) .
\label{eqn:CBBoverCB}
\end{eqnarray}

Since the static antiquarks are separated along the $z$ axis, the symmetry (even in the continuum) is no longer spherical. The quantum numbers are the following: \\
$\phantom{xxx} \bullet \ \ $ $|j_z|$: absolute value of the $z$ component of the total angular momentum of the light quarks \\ $\phantom{xxx \bullet} \ \ $ and gluons. \\
$\phantom{xxx} \bullet \ \ $ $P$: parity. \\
$\phantom{xxx} \bullet \ \ $ $P_x$: behaviour under reflection along the $x$ axis. \\
These quantum numbers can be related to the $4 \times 4$ matrix $\Gamma$ appearing in the interpolating operators (\ref{eqn:creation_operator_bbud}) (see Table~2 in Ref.\ \cite{Bicudo:2015kna}, where $SS$, $SP$ and $PP$ indicate the asymptotic values $2 m_B$, $m_B + m_{B_0^\ast}$ and $2 m_{B_0^\ast}$). In the following we denote the potentials $V^{q^{(1)} q^{(2)},\Gamma}_{BB}(\f{r})$ also by $V^{I;|j_z|,P,P_x}_{BB}(\f{r})$.

Using the Fierz identity, the interpolating operators (\ref{eqn:creation_operator_bbud}) can also be related to $j$, the absolute value of the total angular momentum of the light degrees of freedom of an antiheavy-antiheavy-light-light tetraquark (see Table~5 of Ref.\ \cite{Bicudo:2015kna}). $\Gamma \in \{ \gamma_5 , \gamma_0 \gamma_5 , 1 , \gamma_0 \}$ correspond to $j = 0$, $\Gamma \in \{ \gamma_k , \gamma_0 \gamma_k , \gamma_k \gamma_5 , \gamma_0 \gamma_k \gamma_5\}$ ($k = 1,2,3$) to $j = 1$.

% **********
% **********
% **********
\begin{figure}
	\newcommand{\scale}{0.47}
	\centering
	\includegraphics[width=\scale\textwidth]{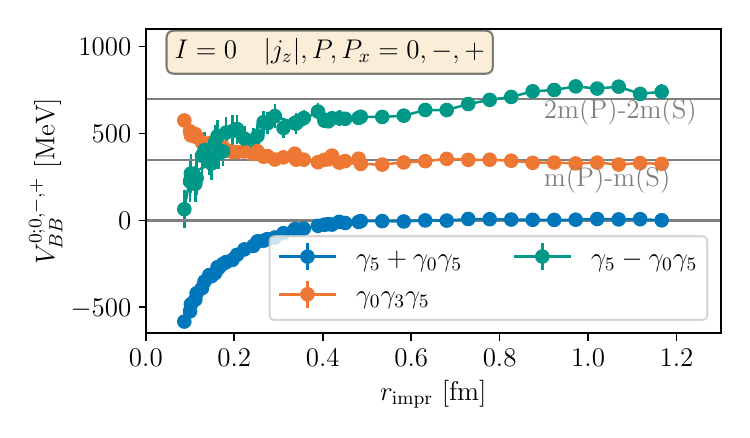}
	\includegraphics[width=\scale\textwidth]{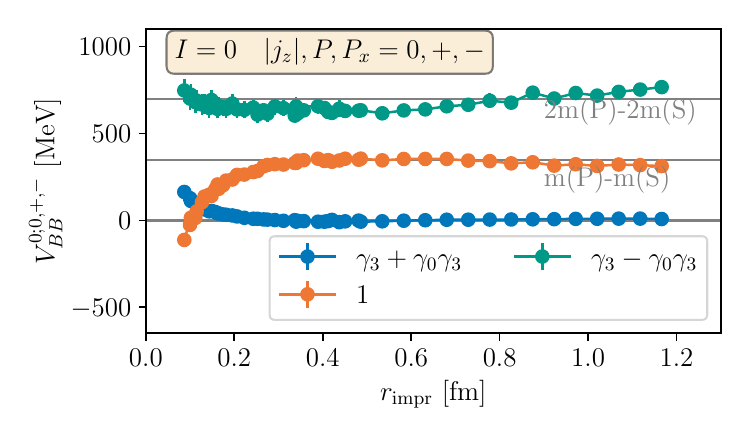}
	\includegraphics[width=\scale\textwidth]{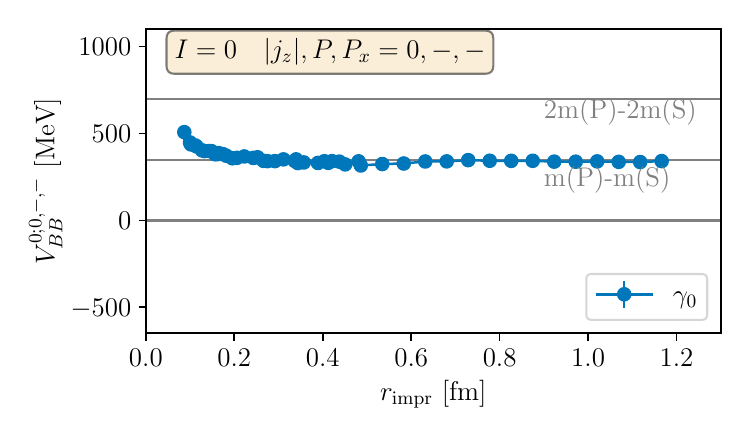}
	\includegraphics[width=\scale\textwidth]{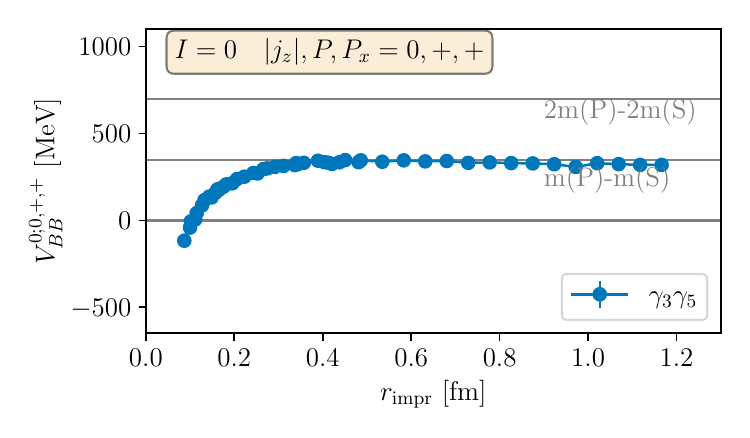}
	\includegraphics[width=\scale\textwidth]{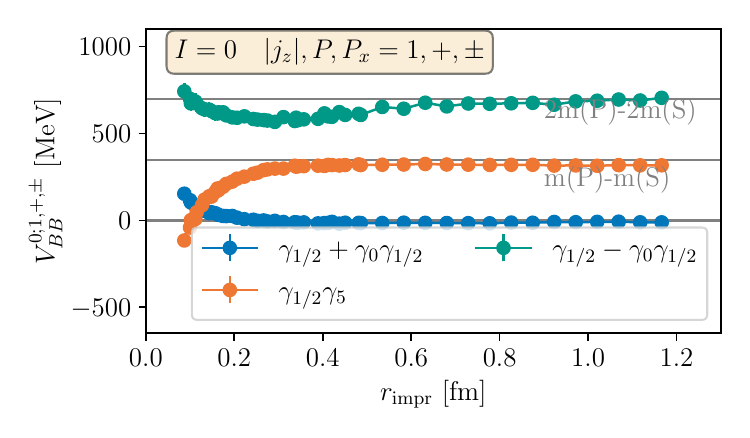}
	\includegraphics[width=\scale\textwidth]{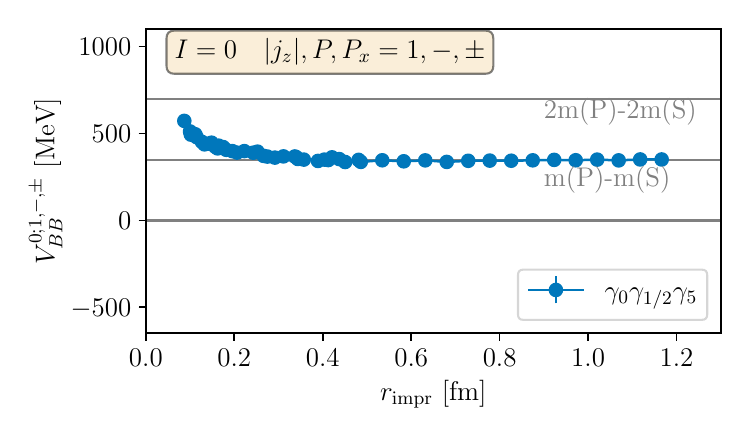}
\caption{$\bar Q \bar Q q q$ potentials for $qq = (ud - du) / \sqrt{2}$, i.e.\ $I = 0$, ensemble N6. $m(S) \equiv m_B \equiv m_{B^\ast}$ denotes the mass of the lightest static-light meson with parity $-$, whereas $m(P) \equiv B_0^\ast \equiv B_1^\ast$ denotes the mass of its parity partner. \label{fig:allI0}}
\end{figure}

% **********
% **********
% **********
\begin{figure}
	\newcommand{\scale}{0.47}
	\centering
	\includegraphics[width=\scale\textwidth]{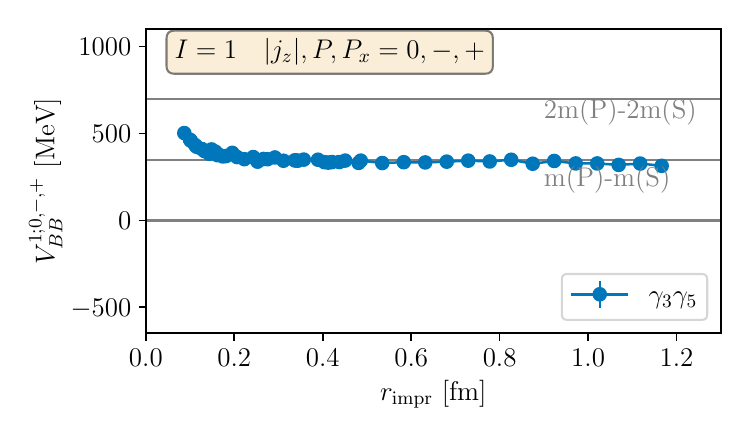}
	\includegraphics[width=\scale\textwidth]{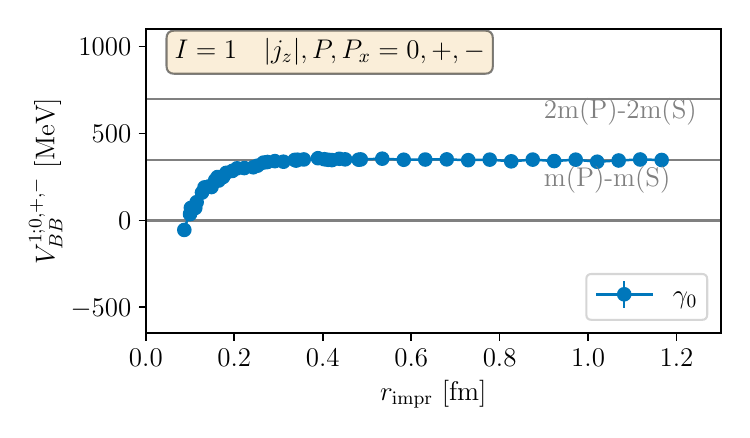}
	\includegraphics[width=\scale\textwidth]{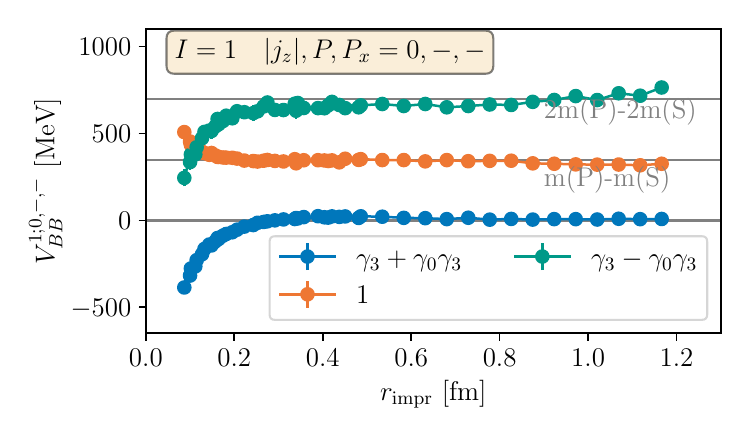}
	\includegraphics[width=\scale\textwidth]{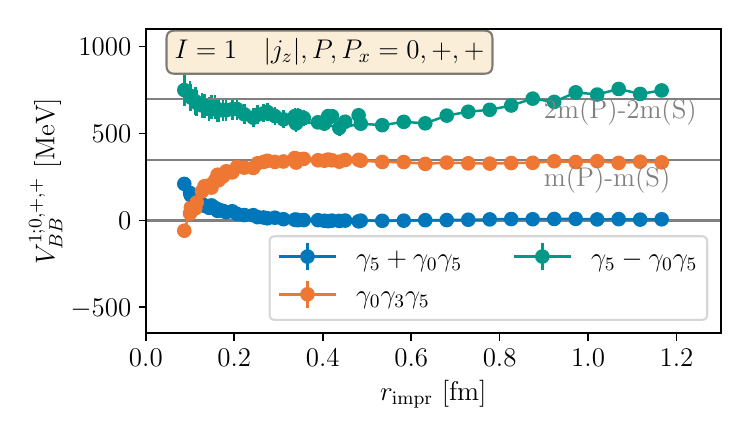}
	\includegraphics[width=\scale\textwidth]{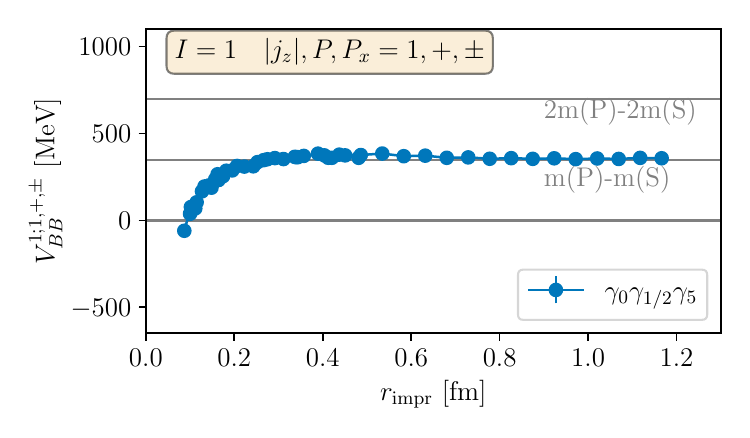}
	\includegraphics[width=\scale\textwidth]{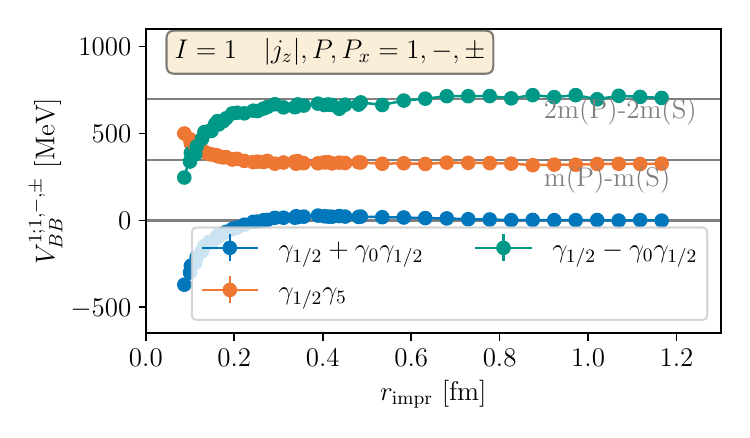}
\caption{$\bar Q \bar Q q q$ potentials for $qq \in \{ uu , (ud + du) / \sqrt{2}, dd \}$, i.e.\ $I = 1$, ensemble N6.  $m(S) \equiv m_B \equiv m_{B^\ast}$ denotes the mass of the lightest static-light meson with parity $-$, whereas $m(P) \equiv B_0^\ast \equiv B_1^\ast$ denotes the mass of its parity partner. \label{fig:allI1}}
\end{figure}

We show our results for $I = 0$ and $I = 1$, obtained on ensemble N6, in Fig.\ \ref{fig:allI0} and Fig.\ \ref{fig:allI1}, respectively (analogous plots for $I = 1/2$ and strangeness $-1$ are qualitatively similar and not shown, because of page limitations). Note that we use correlation functions of an operator (\ref{eqn:creation_operator_bbud}) with itself, but have not yet computed correlation matrices. Our excited potentials might, thus, be contaminated by lower potentials with the same quantum numbers and even for some of the groundstate potentials it is currently not fully clear, whether the limit $t \rightarrow \infty$ indicated in Eq.\ (\ref{eqn:CBBoverCB}) has been reached within statistical errors.

%FFFFFFFFFFFFFFFFFFFFFFFFFFFFFFFFFFFFFFFFFFFFFFFFFFFFFFFFFFFFFFFFFFFFFFFFFFFFFFFFFFFFFFFFFFFFFFFFFF
%FFFFFFFFFFFFFFFFFFFFFFFFFFFFFFFFFFFFFFFFFFFFFFFFFFFFFFFFFFFFFFFFFFFFFFFFFFFFFFFFFFFFFFFFFFFFFFFFFF
%FFFFFFFFFFFFFFFFFFFFFFFFFFFFFFFFFFFFFFFFFFFFFFFFFFFFFFFFFFFFFFFFFFFFFFFFFFFFFFFFFFFFFFFFFFFFFFFFFF

\section{Discussion of our results}

\begin{figure}
	\centering
	\includegraphics[width=1.05\textwidth]{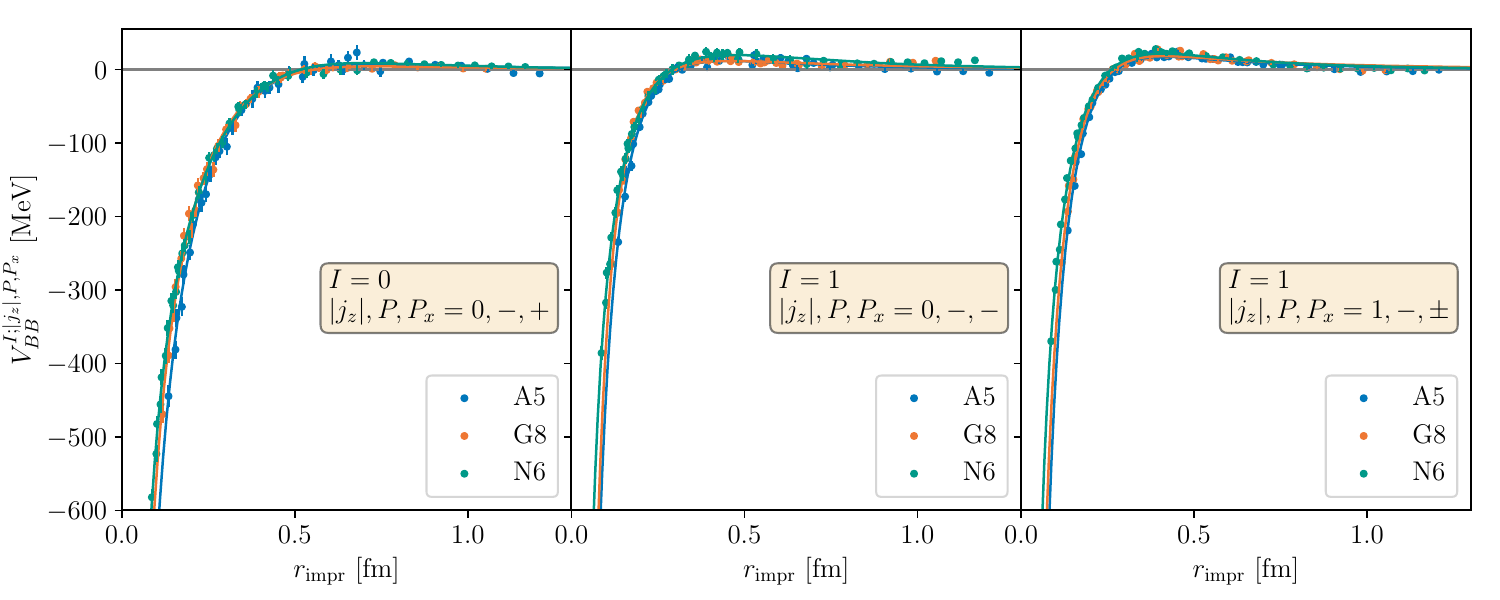}\\
	\includegraphics[width=1.05\textwidth]{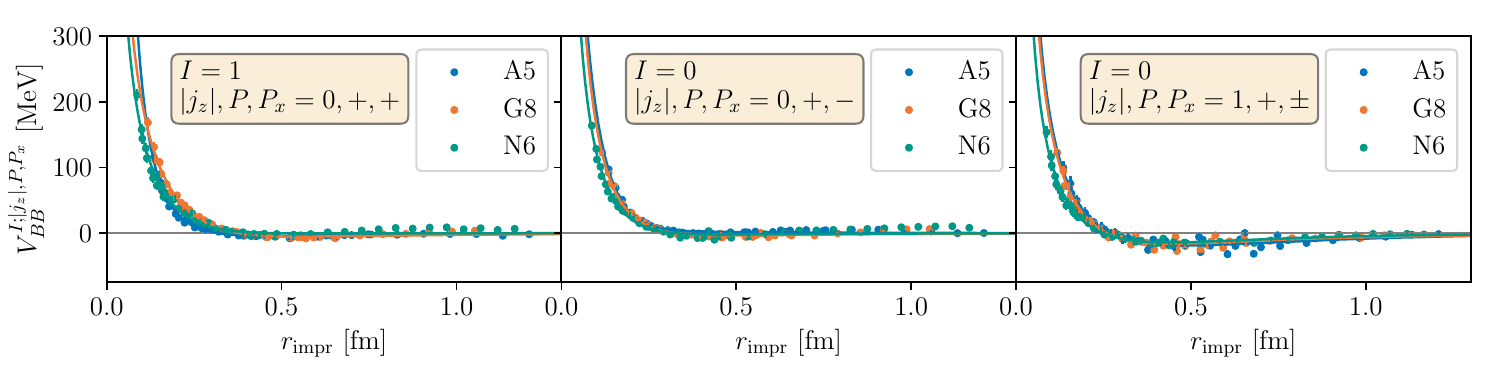}
\caption{$\bar Q \bar Q q q$ potentials with $I = 0$ as well as $I = 1$ and the lowest possible asymptotic value, ensembles A5, G8 and N6.\label{fig:QQll}}
\end{figure}

We now focus mostly on Fig.\ \ref{fig:QQll}, where we show all $I = 0$ and $I = 1$ attractive and repulsive potentials, which have the lowest possible asymptotic value of two times the mass of the lightest static-light meson, i.e.\ $2 m_B$. Analogous plots for $I = 1/2$ and strangeness $-1$ are qualitatively similar and not shown, because of page limitations.

% **********
% **********
% **********

% **********
% **********
% **********
\iffalse
\begin{figure}
	\centering
	\includegraphics[width=1.05\textwidth]{figures/BBs_att.pdf} \\
	\includegraphics[width=1.05\textwidth]{figures/BBs_rep.pdf}
\caption{$\bar Q \bar Q q q$ potentials with $I = 1/2$ and strangeness $-1$ and the lowest possible asymptotic value, ensemble N6. The gray data points and fits represent corresponding results for $I = 0$ and $I = 1$ already shown in Fig.\ \ref{fig:QQll}.\label{fig:QQus}}
\end{figure}
\fi

\begin{itemize}
\item[(i)] 
At small $\bar Q \bar Q$ separations we expect from QCD a one gluon exchange potential. The leading term is $c \alpha_s / r$ with the color factor $c = \boldsymbol \lambda^1 \cdot \boldsymbol \lambda^2 / \text{Tr}(\sbf \lambda \cdot \sbf \lambda)$ ($\boldsymbol \lambda^1$ and $\boldsymbol \lambda^2$ are Gell-Mann matrices acting on the color components of the two heavy antiquarks). $c = -1/2$, if $\bar Q \bar Q$ is in a color triplet, and $c = +1/4$, if $\bar Q \bar Q$ is in a color antisextet.

The potentials in the top row of Fig.\ \ref{fig:QQll} have $I = j$, in the bottom row $I \neq j$. Because of the Pauli principle, the static quarks must be in an antisymmetric color triplet in the top row and in a symmetric color antisextet in the bottom row. This explains, why the potentials in the top row are attractive at small separations, while those in the bottom row are repulsive. The argument can be generalized to potentials with higher asymptotic values in a straightforward way (see Ref.\ \cite{Bicudo:2015kna}).

% -----

\item[(ii)] 
The interaction of the light quark spins also contributes to the potentials. It is known from the hadron spectrum that the dominant contribution is the hyperfine potential. While quark models do not agree on the $r$ dependence of this potential, it is proportional to $-c \boldsymbol \sigma^1 \cdot \boldsymbol \sigma^2$ ($\boldsymbol \sigma^1$ and $\boldsymbol \sigma^2$ are Pauli matrices acting on the spin components of the two light quarks). $-\boldsymbol \sigma^1 \cdot \boldsymbol \sigma^2 \equiv +3$ for $j = 0$ and $-\boldsymbol \sigma^1 \cdot \boldsymbol \sigma^2 \equiv -1$ for $j = 1$.
This explains, why the potentials in the left column of Fig.\ \ref{fig:QQll}, which have $j = 0$, are stronger attractive/repulsive than those in the center and right columns, which have $j = 1$.

% -----

\item[(iii)] 
When increasing the $\bar Q \bar Q$ separation to $\approx 0.25 \, \text{fm}$, a pair of weakly overlapping static-light mesons will form. The potentials are then screened proportional to the tail of the static-light wave-function. This is typically an exponential suppression $\propto \exp(-r^p)$ with exponent $p$ in the range $1.5 , \ldots , 2.0$. This expectation is consistent with our potential data.

% -----

\item[(iv)] 
In the limit $r \rightarrow 0$ our interpolating operators (\ref{eqn:creation_operator_bbud}) become equivalent to interpolating operators for static-light baryons. Thus, the difference between two ground state potentials, both $V^{0;0,-,+}_{BB}(\f{r}) - V^{0;0,-,-}_{BB}(\f{r})$ and $V^{0;0,-,+}_{BB}(\f{r}) - V^{0;1,-,\pm}_{BB}(\f{r})$, should approach the mass difference of the static light baryons with quantum numbers $j^\mathcal{P} = 0^+$ and $j^\mathcal{P} = 1^-$ \cite{Wagner:2011fs} or, equivalently, the mass difference of a ``good'' and a ``bad'' diquark \cite{Francis:2021vrr}. In Fig.\ \ref{fig:goodbad} we show the difference $V^{0;0,-,+}_{BB}(\f{r}) - V^{0;0,-,-}_{BB}(\f{r})$ and find $\approx -200 \, \text{MeV}$ in the limit $r \rightarrow 0$, which is consistent with the static-light baryon and diquark masses from Refs.\ \cite{Wagner:2011fs,Francis:2021vrr}.

% **********
% **********
% **********
\begin{figure}
	\centering
	\includegraphics[width = \textwidth]{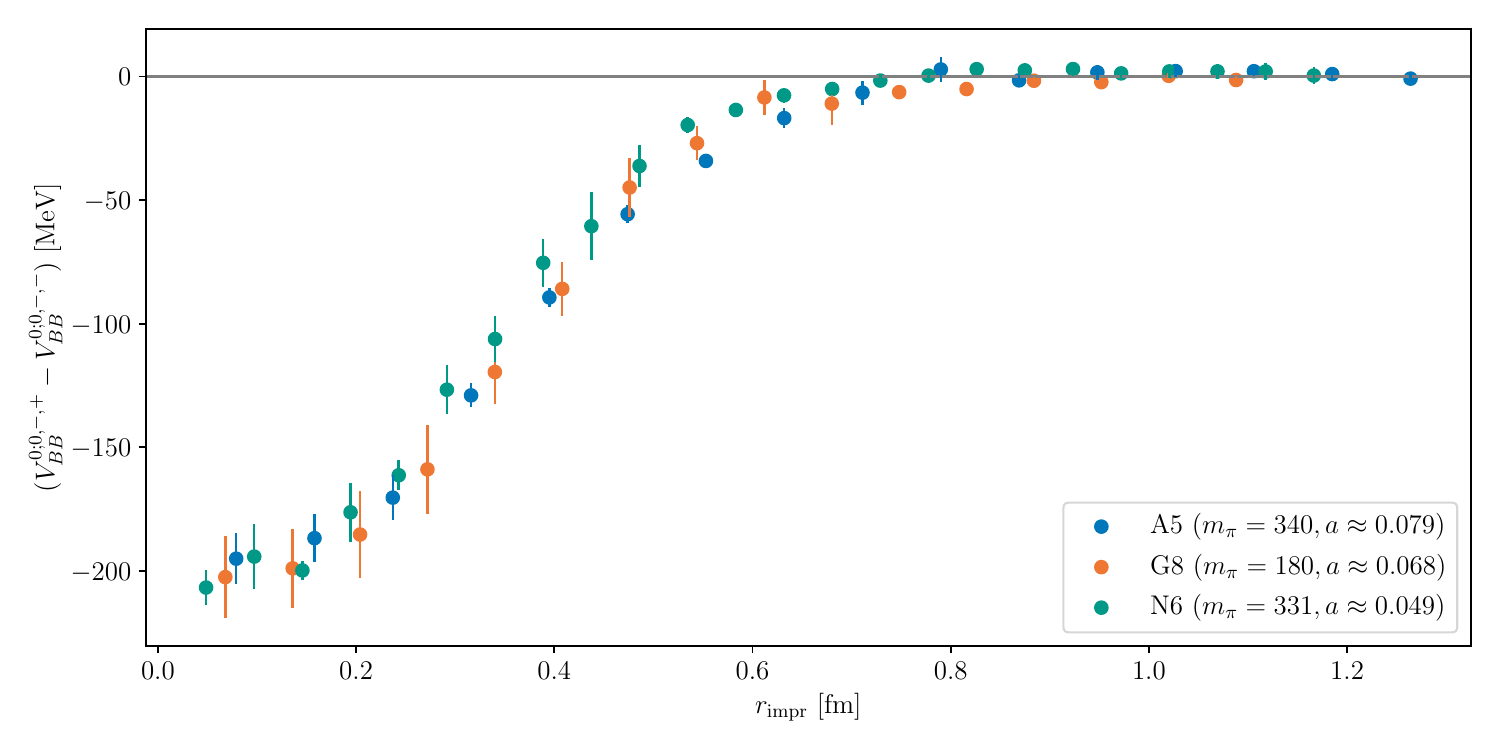}
	\caption{Difference between two ground state potentials, $V^{0;0,-,+}_{BB}(\f{r}) - V^{0;0,-,-}_{BB}(\f{r})$. In the limit $r \rightarrow 0$ this energy difference approaches the mass difference of two static-light baryons or, equvalently, a ``good'' and a ``bad'' diquark.\label{fig:goodbad}}
\end{figure}

% -----

\item[(v)] 
At intermediate $\bar Q \bar Q$ separations $0.25 \, \text{fm} \ltapprox r \ltapprox 0.75 \, \text{fm}$ we obtain for the first time clearly visible bumps for some of the potentials (in Fig.\ \ref{fig:QQll} top row, center and right plot as well as bottom row, right plot), i.e.\ the potentials have a different sign at intermediate separations than at small separations. These bumps may have various non-perturbative contributions: the energy difference in the limit $r \rightarrow 0$ discussed in the previous item (iv), the flip-flop between a tetraquark string and two meson strings \cite{Bicudo:2017usw} as well as meson exchange potentials might all be expected. Note that in the remaining three cases  (top row, left plot as well as bottom row, left and center plot) there might be similar effects leading to  ``bumps'' with the same sign as the short range potential, which are less clearly visible.

% -----

\item[(vi)] 
At large $\bar Q \bar Q$ separations the OPE potential should dominate, which is given by \cite{Beane:2001bc}
\be
{ {m_\pi}^2 {g_A}^2 \over 24 \pi {f_\pi}^2 }
(\sbf \tau_1 \cdot \sbf  \tau_2) \bigg(\Big(3 (\sbf \sigma_1 \cdot \hat{\sbf r}) (\sbf \sigma_2 \cdot \hat{\sbf r}) - \sbf \sigma_1 \cdot \sbf \sigma_2\Big)
\bigg( 1 + {3 \over m_\pi r} + {3 \over (m_\pi r)^2}\bigg) + \sbf \sigma_1 \cdot \sbf \sigma_2\bigg) 
 {e^{- m_\pi r} \over r} .
\ee

Note that it includes a hyperfine contribution proportional to $(\sbf \tau_1 \cdot \sbf \tau_2) (\sbf \sigma_1 \cdot \sbf \sigma_2)$, which, due to the Pauli principle, has the opposite sign as the sort range Coulomb potential discussed in item (i) and, thus, generates a bump at large separations, which might, however, be rather small and difficult to identify.

The main signature of OPE might be the tensor part proportional to $(\sbf \tau_1 \cdot \sbf \tau_2) (\sbf \sigma_1 \cdot \hat{\sbf r}) (\sbf \sigma_2 \cdot \hat{\sbf r})$. In our case $\hat{\sbf r} = \sbf e_z$ and, thus, $(\sbf \sigma_1 \cdot \hat{\sbf r}) (\sbf \sigma_2 \cdot \hat{\sbf r}) = -1$ for $|j_z| = 0$ and $(\sbf \sigma_1 \cdot \hat{\sbf r}) (\sbf \sigma_2 \cdot \hat{\sbf r}) = +1$ for $|j_z| = 1$. Because of this, in Fig.\ \ref{fig:QQll} the tensor interaction is expected to shift the potentials in the center, which have $|j_z| = 0$, in the opposite direction than the potentials at the right, which have $|j_z| =1$. Moreover, notice the center and right plots in the top row correspond to $I = 1$ ($\rightarrow \sbf \tau_1 \cdot \sbf \tau_2 \equiv +1$), while the center and right plots in the bottom row correspond to $I = 0$ ($\rightarrow \sbf \tau_1 \cdot \sbf \tau_2 \equiv -3$). Consequently, observing $(V^{0;1,+,\pm}_{BB}(\f{r}) - V^{0;0,+,-}_{BB}(\f{r})) / (V^{1;1,-,\pm}_{BB}(\f{r}) - V^{1;0,-,-}_{BB}(\f{r})) \approx -3$ at large $\bar Q \bar Q$ separations could be an indication for OPE. Even though statistical errors are large, our data leads to ratios in reasonable agreement with $-3$.
\end{itemize}

%FFFFFFFFFFFFFFFFFFFFFFFFFFFFFFFFFFFFFFFFFFFFFFFFFFFFFFFFFFFFFFFFFFFFFFFFFFFFFFFFFFFFFFFFFFFFFFFFFF
%FFFFFFFFFFFFFFFFFFFFFFFFFFFFFFFFFFFFFFFFFFFFFFFFFFFFFFFFFFFFFFFFFFFFFFFFFFFFFFFFFFFFFFFFFFFFFFFFFF
%FFFFFFFFFFFFFFFFFFFFFFFFFFFFFFFFFFFFFFFFFFFFFFFFFFFFFFFFFFFFFFFFFFFFFFFFFFFFFFFFFFFFFFFFFFFFFFFFFF

\section*{Acknowledgements}

We are grateful to the Coordinated Lattice Simulations (CLS) effort for the access to the $N_f = 2$ gauge link configurations.

P.B.\ acknowledges support by CeFEMA, an IST research unit whose activities are partially funded by the Fundação para a Ciência e a Tecnologia -- FCT contract UI/DB/04540/2020 for R\&D Units.
M.K.M.\ would like to express a special thanks to the Mainz Institute for Theoretical Physics (MITP) of the DFG Cluster of Excellence PRISMA+ (Project ID 39083149), for its hospitality and support.
L.M.\ and M.W.\ acknowledge support by the Deutsche Forschungsgemeinschaft (DFG, German Research Foundation) -- project number 457742095.
M.W.\ acknowledges support by the Heisenberg Programme of the Deutsche Forschungsgemeinschaft (DFG, German Research Foundation) -- project number 399217702.

We acknowledge access to Piz Daint at the Swiss National Supercomputing Centre, Switzerland under the ETHZ’s share with the project ID eth8 and s1193.
Calculations on the GOETHE-NHR and on the FUCHS-CSC  high-performance computers of the Frankfurt University were conducted for this research. We would like to thank HPC-Hessen, funded by the State Ministry of Higher Education, Research and the Arts, for programming advice.

%FFFFFFFFFFFFFFFFFFFFFFFFFFFFFFFFFFFFFFFFFFFFFFFFFFFFFFFFFFFFFFFFFFFFFFFFFFFFFFFFFFFFFFFFFFFFFFFFFF
%FFFFFFFFFFFFFFFFFFFFFFFFFFFFFFFFFFFFFFFFFFFFFFFFFFFFFFFFFFFFFFFFFFFFFFFFFFFFFFFFFFFFFFFFFFFFFFFFFF
%FFFFFFFFFFFFFFFFFFFFFFFFFFFFFFFFFFFFFFFFFFFFFFFFFFFFFFFFFFFFFFFFFFFFFFFFFFFFFFFFFFFFFFFFFFFFFFFFFF

\bibliographystyle{JHEP}
\bibliography{proceedingsLattice2024.bib}

%%%%%%%%%%%%%%%%%%%%%%%%%%%%%%%%%%%%%%%%%%%%%%%%%%%%%%%%%%%%%%%%%%%%%%%%%%%%%
\end{document}